\documentclass[12pt]{article}
\usepackage{amssymb}
\usepackage{amsmath}
\usepackage{amsfonts}
\usepackage{scicite}
\usepackage{times}
\usepackage{amsthm}

\newcounter{lastnote}

\newtheorem{theorem}{Theorem}

\newtheorem{definition}[theorem]{Definition}

\newtheorem{lemma}[theorem]{Lemma}

\newtheorem{proposition}[theorem]{Proposition}

\newtheorem*{th0}{Theorem 0}
\input{tcilatex}

\begin{document}

\title{The Housing Problem and Revealed Preference Theory: Duality and an
Application}
\author{Ivar Ekeland \\
(Universit\'{e} Paris-Dauphine) \and Alfred Galichon \\
(Ecole polytechnique)}
\date{{\small August 28, 2012\thanks{%
Correspondence: Economics Department, Ecole polytechnique, 91128 Palaiseau,
France. E-mail: alfred.galichon@polytechnique.edu. Support from Chaire
EDF-Calyon \textquotedblleft Finance and D\'{e}veloppement
Durable\textquotedblright\ \ and FiME, Laboratoire de Finance des March\'{e}%
s de l'Energie (www.fime-lab.org) is gratefully acknowledged by both
authors, and from Chaire Axa \textquotedblleft Assurance et Risques
Majeurs,\textquotedblright\ Chaire FDIR \textquotedblleft Socially
Responsible Investments and Value Creation\textquotedblright\ by the second
author. The authors thank John Geanakoplos for communicating them his paper
\textquotedblleft Afriat from MaxMin\textquotedblright . This paper has
benefited from insightful comments by Don Brown, Federico Echenique, Fran%
\c{c}oise Forges, Peter Hammond, Vincent Iehl\'{e} and Enrico Minelli.

This is a pre-print of an article published in \textit{Economic Theory}. The final authenticated version is available online at: https://doi.org/10.1007/s00199-012-0719-x}.}}
\maketitle

\begin{abstract}
This paper exhibits a duality between the theory of Revealed Preference of
Afriat and the housing allocation problem of Shapley and Scarf. In
particular, it is shown that Afriat's theorem can be interpreted as a second
welfare theorem in the housing problem. Using this duality, the revealed
preference problem is connected to an optimal assignment problem, and a
geometrical characterization of the rationalizability of experiment data is
given. This allows in turn to give new indices of rationalizability of the
data, and to define weaker notions of rationalizability, in the spirit of
Afriat's efficiency index.
\end{abstract}

\noindent

{\footnotesize \textbf{JEL codes}: primary D11, secondary C60, C78. \vskip%
50pt Keywords: revealed preferences, Afriat's theorem, optimal assignments,
indivisible allocations.}

\newpage



\section{Introduction}

The \textquotedblleft Revealed Preference\textquotedblright\ problem and the
\textquotedblleft Housing problem\textquotedblright\ are two classical
problems in Economic Theory with for both a distinguished, but separate
tradition. This paper is about connecting them, and exploit the connection
to obtain new results on Revealed Preference.

\bigskip

\begin{itemize}
\item The Revealed Preference (RP) problem, posed by Samuelson at the end of
the 1930's, was solved by Houthakker in 1950, and was given an operational
solution by Sidney Afriat in 1967. This classical problem asks whether,
given the observation of $n$ consumptions baskets and corresponding prices,
one rationalize these consumptions as the consumption of a single consumer
facing different prices.

\item The \textquotedblleft housing problem\textquotedblright\ was
investigated in 1974 by Shapley and Scarf. Given an initial allocation of $n$
houses to $n$ individuals, and assuming individuals form preferences over
houses and can trade houses, what is the core of the corresponding game? It
is assumed that houses form no preferences over owners (in sharp contrast to
the \textquotedblleft Stable matching\textquotedblright\ problem of Gale and
Shapley). In this setting, they showed non-emptiness of the core, as well as
an algorithm to arrive to a core allocation: the method of \textquotedblleft
top-trading cycles\textquotedblright , attributed to David Gale.
\end{itemize}

\bigskip

Although both problems have generated two well established and distinct
literatures, it will turn out that these problems are in fact dual in a
precise sense. As we shall argue, the traditional expenditure/utility
duality from consumer theory extends to the setting of revealed preference,
and it is possible to show that the problem of Pareto efficiency in the
housing problem and the issue of rationalizability of consumer data are both
applications of a basic mathematical result characterizing cyclically
consistent matrices proven by Fostel, Scarf and Todd (2004), generalizing
Afriat's theorem. In particular, we show that \emph{Afriat's theorem can be
interpreted as a second welfare theorem in the housing problem}. Once we
have established equivalence of both problems, we shall use the
aforementioned results to give new characterization of both problems in
terms of an optimal allocation problem. This will give us a very simple and
intuitive characterization of the efficient outcomes in the indivisible
allocation problem.

According to standard revealed preference theory, the data are not
rationalizable when one can exhibit a preference cycle based on the direct
revealed preference relation. A stronger violation of rationalizability
occurs when there are subsets of the data where cycles exist. An even
stronger departure from rationalizability occurs when all observations are
part of such a cycle. We shall characterize these three situations in terms
of indexes which will measure how strong the violations of rationalizability
are. In the last part, we shall last use these results to provide a
geometric interpretation of revealed preference theory, by showing that the
problem of determining whether data can be rationalized or not can be recast
as the problem of determining whether a particular point is on some part of
the boundary of some convex body.

The connection between the housing problem and the revealed preference
problem yields some insights for both theories. The important econometrics
literature on revealed preference contains a wealth of tools to measure the
intensity of departure from rationalizability. These tools may be
successfully applied to the \textquotedblleft housing
problem\textquotedblright\ and its more applied versions such as the
\textquotedblleft kidney problem\textquotedblright , which deals with
assigning organ transfers to patients. These tools may design indices to
measure how far from Pareto optimality an actual assignment is, or what are
the costs of a given constraint in terms of welfare. Conversely, the
connection between the two problems sheds light on the role of matching
theory into revealed preference theory. This leads to interesting research
avenues, as matching theory has recently received a lot of attention, both
from the empirical and computational point of views.

\bigskip

The literature on revealed preference in consumer demand traces back to
Samuelson (1938), who formulated the problem and left it open. It was solved
by Afriat (1967) using nontrivial combinatorial techniques. Diewert (1973)
provided a Linear Programming proof and Varian (1982) an algorithmic
solution. Fostel, Scarf and Todd (2004) and Chung-Piaw and Vohra (2003)
provided alternative proofs. Matzkin (1991) and Forges and Minelli (2009)
extended the theory to nonlinear budget constraints. Geanakoplos (2006)
gives a proof of Afriat's theorem using a minmax theorem. Although min-max
formulations will appear in our paper, these are distinct from Geanakoplos',
as it will be explained below. The setting was recently extended to
collective models of household consumptions by Cherchye, De Rock and
Vermeulen in a series of paper (see e.g. 2007, 2010). A very recent paper by
Quah (2012) extends Afriat's Theorem in another direction n order to
characterize data sets compatible with weak separability without concavity
assumptions. The literature on the indivisible allocation problem was
initiated by Shapley and Scarf (1974), who formulated as the
\textquotedblleft housing problem\textquotedblright\ and gave an abstract
characterization of the core. Roth et al (2004) study a related
\textquotedblleft kidney problem\textquotedblright\ and investigate
mechanism design aspect. Finally, recent literature has extended revealed
preference theory to classes of matching problems. Galichon and Salani\'{e}
(2010) investigate the problem of revealed preferences in a matching game
with transferable utility, with and without unobserved heterogeneity.
Echenique et al. (2011) investigate the problem of revealed preferences in
games with and without transferable utility, and without unobserved
heterogeneity.

\bigskip

This paper will be organized as follows. We first study Pareto efficient
allocations, and show the connection with the generalized theory of revealed
preferences. We deduce a novel characterization of data rationalizability in
terms of an optimal matching problem, and a geometric interpretation of
revealed preference in terms of convex geometry. We then move on to
providing indexes of increasingly weaker forms of rationalizability.

\bigskip

\section{Revealed preference and the housing problem}

\subsection{The Generalized Afriat's theorem}

Assume as in Forges and Minelli (2009) that consumer has budget constraint $%
g_{i}\left( x\right) \leq 0$ in experiment $i$ where $x$ is an $L$%
-dimensional bundle of goods, and that bundle $x_{i}$ is chosen where $%
g_{i}\left( x_{i}\right) =0$. This is a nonlinear generalization of Afriat
(1969), in which $g_{i}\left( x_{j}\right) =x_{j}\cdot p_{i}-x_{i}\cdot
p_{i} $, where $p_{i}$ is the price vector of the bundle of goods in
experiment $i$. Forges and Minelli ask whether there exists a utility
function $v\left( x\right) $ with appropriate properties such that 
\begin{equation*}
x_{i}\in \func{argmax}_{x}\left\{ v\left( x\right) :g_{i}\left( x\right)
\leq 0\right\}
\end{equation*}%
in which case $v$ is said to \emph{rationalize} the data.

\bigskip

As in the original Afriat's paper, the first part of their proof
necessitates the existence of a utility level $v_{j}$ associated to bundle $%
j $ such that 
\begin{equation}
g_{i}\left( x_{j}\right) <0\text{ implies }v_{j}-v_{i}<0.
\label{utilityLevels}
\end{equation}

Once the number $v_{i}$'s have been determined, the second part of their
proof consists in constructing a locally nonsatiated and continuous utility
function $v\left( x\right) $ such that $v\left( x_{j}\right) =v_{j}$. The
preference induced by $v$ are said to rationalize the data in the sense that 
\begin{equation*}
i\in \func{argmax}_{j}\left\{ v_{j}:g_{i}\left( x_{j}\right) \leq 0\right\} .
\end{equation*}

\bigskip

The existence of real numbers $v_{i}$ such that (\ref{utilityLevels}) holds
is non-trivial and it turns out to be equivalent to a property of matrix $%
R_{ij}=g_{i}\left( x_{j}\right) $ called \textquotedblleft cyclical
consistency\textquotedblright . This is done by appealing to an extension of
Afriat's original theorem beyond linear budget sets, proved by Fostel, Scarf
and Todd (2004). This result can be slightly reformulated as follows:

\bigskip

\begin{th0}[Afriat's theorem]
\label{ThmZero}The following conditions are equivalent:\bigskip

(i) The matrix $R_{ij}$ satisfies \textbf{\textquotedblleft cyclical
consistency\textquotedblright }: for any cycle $i_{1},...,i_{p+1}=i_{1}$,%
\begin{equation}
\forall k,~R_{i_{k}i_{k+1}}\leq 0\text{ implies }\forall
k,~R_{i_{k}i_{k+1}}=0,  \label{cyclicalConsistency}
\end{equation}

(ii) There exist numbers $\left( v_{i},\lambda _{i}\right) $, $\lambda
_{i}>0 $, such that 
\begin{equation*}
v_{j}-v_{i}\leq \lambda _{i}R_{ij},
\end{equation*}

(iii) There exist numbers $v_{i}$ such that%
\begin{eqnarray*}
R_{ij} &\leq &0\text{ implies }v_{j}-v_{i}\leq 0,\text{ and} \\
R_{ij} &<&0\text{ implies }v_{j}-v_{i}<0.
\end{eqnarray*}
\end{th0}

\bigskip

Before we give the proof of the result, let us briefly comment on it.
Although we call it \textquotedblleft Afriat's theorem\textquotedblright ,\
this result is in some sense a fundamental lemma upon which Afriat's
analysis rests. In the original Afriat's work and in the subsequent
literature, notably Forges and Minelli (2009), a full description of the
budget sets is provided, that is the functions $g_{i}\left( x\right) $
defined on a larger domain than the set of observed consumption bundles $%
\left\{ x_{1},...,x_{n}\right\} $. Here, the result that we give only
depends on the value $g_{i}\left( x_{j}\right) $ of the functions $g_{i}$
over the set of the observed consumption bundles. Actually, even only the
signs of these values matter. Here the analysis is intrinsiquely discrete
and the approach relies neither on a particular consumption space, nor on
explicit consumption bundles $x_{j}$.

\bigskip

We now turn to the proof of Theorem 0.

\begin{proof}[Proof of Theorem 0]
(i) implies (ii) is proven in Fostel, Scarf and Todd (2004). (ii)
immediately implies (iii). We now show that (iii) implies (i). Consider a
cycle $i_{1},...,i_{p+1}=i_{1}$, such that%
\begin{equation*}
\forall k,~R_{i_{k}i_{k+1}}\leq 0\text{.}
\end{equation*}

Then by (iii) there exist numbers $v_{i}$ such that%
\begin{eqnarray*}
R_{ij} &\leq &0\text{ implies }v_{j}-v_{i}\leq 0,\text{ and} \\
R_{ij} &<&0\text{ implies }v_{j}-v_{i}<0.
\end{eqnarray*}%
thus one has $v_{i_{k+1}}-v_{i_{k}}\leq 0$ for all $k$, hence all the $%
v_{i_{k}}$ are equal. Assume now that there is a $k$ such that $%
R_{i_{k}i_{k+1}}<0$. Then $v_{i_{k+1}}-v_{i_{k}}<0$, which is a
contradiction. Therefore,
\begin{equation*}
\forall k,~R_{i_{k}i_{k+1}}=0,
\end{equation*}%
which proves the cyclical consistency of matrix $R$, that is (i).
\end{proof}

\bigskip

\subsection{Pareto efficient allocation of indivisible goods}

We now turn to the problem of allocation of indivisible goods, which was
initially studied by Shapley and Scarf (1974). Consider $n$ indivisible
goods (eg. houses) $j=1,...,n$ to be allocated to $n$ individuals. Cost of
allocating (eg. transportation cost) house $j$ to individual $i$ is $c_{ij}$%
. An allocation is a permutation $\sigma $ of the set $\left\{
1,..,n\right\} $ such that individual $i$ gets house $j=\sigma \left(
i\right) $. Let $S$ be the set of such permutations. We assume for the
moment that the initial allocation is given by the identity permutation:
good $i$ is allocated to individual $i$. The problem here is to decide
whether this allocation is efficient in a Pareto sense.

\bigskip

If there are two individuals, say $i_{1}$ and $i_{2}$ that would both
benefit from swapping houses (strictly for at least one), then this
allocation is not efficient, as the swap would improve on the welfare of
both individuals. Thus if allocation is efficient, then inequalities $%
c_{i_{1}i_{2}}\leq c_{i_{1}i_{1}}$ and $c_{i_{2}i_{1}}\leq c_{i_{2}i_{2}}$
cannot hold simultaneously unless they are both equalities. More generally,
Pareto efficiency rules out the existence of exchange rings whose members
would benefit (strictly for at least one) from a circular trade. In fact, we
shall argue that \emph{this problem is dual to the problem of Generalized
Revealed Preferences}.

\bigskip

\subsection{A dual interpretation of revealed preference}

Let us now formalize the notion of efficient allocation in the previous
discussion. Allocation $\sigma _{0}\left( i\right) =i$ is Pareto efficient
if and only if for any $\sigma \in S$, inequalities 
\begin{equation*}
c_{i\sigma \left( i\right) }\leq c_{ii}
\end{equation*}%
cannot hold simultaneously unless these are all equalities. By the
decomposition of a permutation into cycles, we see that this definition is
equivalent to the fact that for every \textquotedblleft trading
cycle\textquotedblright\ $i_{1},...,i_{p+1}=i_{1}$, 
\begin{equation*}
\forall k,\text{ }c_{i_{k}i_{k+1}}\leq c_{i_{k}i_{k}}\text{ implies }\forall
k,\text{ }c_{i_{k}i_{k+1}}=c_{i_{k}i_{k}}
\end{equation*}%
that is, introducing $R_{ij}=c_{ij}-c_{ii}$,%
\begin{equation*}
\forall k,\text{ }R_{i_{k}i_{k+1}}\leq 0\text{ implies }\forall k,\text{ }%
R_{i_{k}i_{k+1}}=0\text{,}
\end{equation*}%
which is to say that allocation is efficient if and only if the matrix $%
R_{ij}$ is cyclically consistent.

\bigskip

By the equivalence between (i) and (ii) in Theorem 0 above, we have the
following statement:

\begin{proposition}
In the housing problem, allocation $\sigma _{0}\left( i\right) =i$ is
efficient if and only if%
\begin{equation}
\exists v_{i}\text{ and }\lambda _{i}>0\text{, }v_{j}-v_{i}\leq \lambda
_{i}R_{ij}.  \tag{PARETO}
\end{equation}
\end{proposition}

\bigskip

Before giving an interpretation of this result, we would like to understand
the link between efficiency and equilibrium in the housing problem. Assume
we start from allocation $\sigma _{0}\left( i\right) =i$, and we let people
trade. Given a price system $\pi $ where $\pi _{i}$ is the price of house $i$%
, we assume that individual $i$ can sell her house for price $\pi _{i}$, and
therefore can afford any house $j$ whose price $\pi _{j}$ is less than $\pi
_{i}$. Therefore, individual $i$'s budget set $B_{i}$ is the set of houses
that sell to a price lower than his'%
\begin{equation*}
B_{i}=\left\{ j:\pi _{j}\leq \pi _{i}\right\}
\end{equation*}

\bigskip

We can now define the notion of a competitive equilibrium in this setting.
Allocation $\sigma \in S$ is an equilibrium supported by price system $\pi $
if for each individual $i$, 1) house $\sigma \left( i\right) $ is weakly
preferred by individual $i$ among all houses that she can afford, while 2)
house $\sigma \left( i\right) $ is strictly preferred by $i$ among all
houses she can strictly afford (i.e., that trade for prices strictly less
than her house $i$). While condition 1) is a necessary requirement,
refinement 2) is natural from a behavioral point of view as it implies that
if $i$ is indifferent between two houses, then she is going to choose the
cheapest house of the two.

\bigskip

In particular, $\sigma _{0}\left( i\right) =i$ is a \emph{No-Trade
equilibrium} if there is a system of prices $\pi $, where $\pi _{j}$ is the
price of house $j$, such that: 1) whenever house $j$ is affordable for
individual $i$, then it is not strictly preferred by $i$ to $i$'s house,
that is $\pi _{j}\leq \pi _{i}$ implies $c_{ij}\geq c_{ii}$, and 2) for any
house $j$ in the strict interior of $i$'s budget set, i.e. that trades for
strictly cheaper than house $i$, then individual $i$ strictly prefers her
own house $i$ to house $j$, that is $\pi _{j}<\pi _{i}$ implies $%
c_{ij}>c_{ii}$.

\bigskip

Hence, $\sigma _{0}\left( i\right) =i$ is a No-Trade equilibrium supported
by prices $\pi $ when conditions (E1) and (E2) below are met, where:

\bigskip

\textbf{(E1)} if house $j$ can be afforded by $i$, then individual $i$ does
not strictly prefer house $j$ to house $i$, that is,%
\begin{equation*}
\pi _{j}\leq \pi _{i}\text{ implies }c_{ij}\geq c_{ii}\text{,}
\end{equation*}%
that is%
\begin{equation*}
\pi _{j}\leq \pi _{i}\text{ implies }R_{ij}\geq 0\text{,}
\end{equation*}%
that is, yet equivalently:%
\begin{equation*}
R_{ij}<0\text{ implies }\pi _{j}>\pi _{i}.
\end{equation*}

\bigskip

\textbf{(E2)} if house $j$ is (strictly) cheaper than house $i$, then
individual $i$ strictly prefers house $i$ to house $j$, that is%
\begin{equation*}
\pi _{j}<\pi _{i}\text{ implies }c_{ij}>c_{ii}\text{,}
\end{equation*}%
that is%
\begin{equation*}
\pi _{j}<\pi _{i}\text{ implies }R_{ij}>0\text{,}
\end{equation*}%
that is, yet equivalently:%
\begin{equation*}
R_{ij}\leq 0\text{ implies }\pi _{j}\geq \pi _{i}.
\end{equation*}

\bigskip

\begin{proposition}
In the housing problem, allocation $\sigma _{0}\left( i\right) =i$ is a
No-Trade equilibrium supported by prices $\pi $ if and only if 
\begin{eqnarray}
R_{ij} &<&0\text{ implies }\pi _{j}>\pi _{i},\text{ and} 
\TCItag{EQUILIBRIUM} \\
R_{ij} &\leq &0\text{ implies }\pi _{j}\geq \pi _{i}.  \notag
\end{eqnarray}
\end{proposition}

\bigskip

But (EQUILIBRIUM) is exactly formulation (iii) of Theorem 0 with $\pi
_{i}=-v_{i}$. By Theorem 0, we know that this statements is equivalent to
statement (PARETO). Hence, we get an interpretation of \emph{the Generalized
Afriat's Theorem as a second welfare theorem}%
\begin{equation*}
\text{(EQUILIBRIUM) }\Longleftrightarrow \text{ (PARETO),}
\end{equation*}%
which we summarize in the following proposition:

\begin{proposition}
In the allocation problem of indivisible goods, Pareto allocations are
no-trade equilibria supported by prices, and conversely, no-trade equilibria
are Pareto efficient.
\end{proposition}

\bigskip

This is a \textquotedblleft dual\textquotedblright\ interpretation of
revealed preference, where $v_{i}$ (utilities in generalized RP theory)
become budgets here, and $c_{ij}$ (budgets in generalized RP theory) become
utilities here. To summarize this duality, let us give the following table:

\begin{center}
\begin{tabular}{l|l|l|}
\cline{2-3}
& {\small Revealed prefs.} & {\small Pareto indiv. allocs.} \\ \hline\hline
\multicolumn{1}{||l|}{\small setting} & \multicolumn{1}{||l|}{\small %
consumer demand} & \multicolumn{1}{||l||}{\small allocation problem} \\ 
\hline\hline
\multicolumn{1}{||l|}{\small budget sets} & \multicolumn{1}{||l|}{$\left\{
j:c_{ij}\leq c_{ii}\right\} $} & \multicolumn{1}{||l||}{$\left\{ -v:-v\leq
-v_{i}\right\} $} \\ \hline\hline
\multicolumn{1}{||l|}{{\small cardinal utilities to }$j$} & 
\multicolumn{1}{||l|}{$v_{j}$} & \multicolumn{1}{||l||}{$-c_{ij}${\small \ }}
\\ \hline\hline
\multicolumn{1}{||l|}{\small \# of consumers} & \multicolumn{1}{||l|}{\small %
one} & \multicolumn{1}{||l||}{${\small n}$, $i\in \left\{ 1,...,n\right\} $}
\\ \hline\hline
\multicolumn{1}{||l|}{\small \# of experiments} & \multicolumn{1}{||l|}{$%
{\small n}$} & \multicolumn{1}{||l||}{\small one} \\ \hline\hline
\multicolumn{1}{||l|}{{\small unit of }$c_{ij}$} & \multicolumn{1}{||l|}%
{\small dollars} & \multicolumn{1}{||l||}{\small utils} \\ \hline\hline
\multicolumn{1}{||l|}{{\small unit of }$v_{i}$} & \multicolumn{1}{||l|}%
{\small utils} & \multicolumn{1}{||l||}{\small dollars} \\ \hline\hline
\multicolumn{1}{||l|}{\small interpretation} & \multicolumn{1}{||l|}{\small %
Afriat's theorem} & \multicolumn{1}{||l||}{\small Welfare theorem} \\ 
\hline\hline
\end{tabular}
\end{center}

The table suggests that there is a single agent facing $n$ choices in
revealed preference theory, while there are $n$ agents facing one choice in
the housing probem. In revealed preference theory, the cardinal utility of
alternative $j$ of the consumer in revealed preference theory is $v_{j}$,
while $c_{ij}$ expresses a budget set. Conversely, in the housing problem, $%
-c_{ij}$ is the cardinal utlity consumer $i$ for alternative $j$, while $%
-v_{j}$ is interpreted as the price of bundle $j$. Hence, the roles of
\textquotedblleft prices\textquotedblright\ and \textquotedblleft
utilities\textquotedblright\ are exchanges in the two problems, which
motivates our contention that Revealed Preference theory and the Housing
problem are \textquotedblleft dual\textquotedblright .

\bigskip 

It is worth remarking that the notion of domination used in our definition
of Pareto efficiency is different from the one used by Shapley and Scarf.
Following a common use, our notion allows for weak domination for some
individual, while Shapley and Scarf require strict domination for everybody.
The two notions are not equivalent in a setting with indivisible goods.
Indeed, using their notion of domination, Shapley and Scarf provide an
example in which a Pareto efficient allocation cannot be sustained as a
competitive allocation

\bigskip

\subsection{A characterization of rationalizability}

The connection of both the revealed preference problem and the housing
problem with assignment problems is clear; we shall now show that there is a
useful connection with the optimal assignment problem (recalled below),
where the sum of the individual utilities is maximized.

Recall that the data are called \emph{rationalizable} there are scalars $%
v_{i}$ such that%
\begin{eqnarray}
R_{ij} &=&g_{i}\left( x_{j}\right) <0\text{ implies }v_{j}<v_{i},\text{ and}
\TCItag{EQUILIBRIUM} \\
R_{ij} &=&g_{i}\left( x_{j}\right) \leq 0\text{ implies }v_{j}\leq v_{i}. 
\notag
\end{eqnarray}%
With this in mind, we have the following novel characterization of
rationalizability of the data, which extends Theorem 0:

\begin{theorem}
\label{ThmCharacterizationGARP}In the revealed preference problem, the data
are rationalizable if and only if there exist weights $\lambda _{i}>0$%
\thinspace\ such that 
\begin{equation}
\min_{\sigma \in S}\sum_{i=1}^{n}\lambda _{i}R_{i\sigma \left( i\right) }=0,
\label{optTransp}
\end{equation}%
that is%
\begin{equation}
\min_{\sigma \in S}\sum_{i=1}^{n}\lambda _{i}c_{i\sigma \left( i\right)
}=\sum_{i=1}^{n}\lambda _{i}c_{ii}.  \label{negishiOpt}
\end{equation}
\end{theorem}

\bigskip

\begin{proof}[Proof of Theorem \protect\ref{ThmCharacterizationGARP}]
Start from (\ref{negishiOpt}): $ ~\exists \lambda
_{i}>0,~\min_{\sigma \in S}\sum_{i=1}^{n}\lambda _{i}R_{i\sigma \left(
i\right) }=0$

$\Longleftrightarrow ~\exists \lambda _{i}>0,~\min_{\sigma \in
S}\sum_{i=1}^{n}\lambda _{i}R_{i\sigma \left( i\right) }$ is reached for $%
\sigma =Id$

\bigskip

This problem is to find the assignment $\sigma \in S$ which minimizes the
utilitarian welfare loss computed as the sum of the individual costs $%
K_{ij}=\lambda _{i}R_{ij}$, setting weight one to each individuals. This
problem is therefore%
\begin{equation*}
\min_{\sigma \in S}\sum_{i=1}^{n}K_{i\sigma \left( i\right) }.
\end{equation*}%
This problem was reformulated as a Linear Programming problem by Dantzig in
the 1930s; see Shapley and Shubik 1971 for a game-theoretic interpretation, and Gretsky, Ostroy and Zame 1992 for the continuous limit),
and by the standard Linear Programming duality of the optimal assignment
problem (Dantzig 1939; Shapley-Shubik 1971; see Ekeland 2010 for a recent review using the theory of Optimal Transportation)%
\begin{equation}
\min_{\sigma \in S}\sum_{i=1}^{n}K_{i\sigma \left( i\right)
}=\max_{u_{i}+v_{j}\leq K_{ij}}\sum_{i=1}^{n}u_{i}+\sum_{j=1}^{n}v_{j},
\label{LPMatching}
\end{equation}%
where $S$ is the set of permutations of $\left\{ 1,...,n\right\} $. It is
well-known that for a $\sigma _{0}\in S$ solution to the optimal assignment
problem, there is a pair $\left( u,v\right) $ solution to the dual problem
such that%
\begin{eqnarray*}
u_{i}+v_{j} &\leq &K_{ij} \\
\text{if }j=\sigma _{0}\left( i\right) \text{, } &&\text{then }%
u_{i}+v_{j}=K_{ij}\text{.}
\end{eqnarray*}

Hence, (\ref{negishiOpt})

$\Longleftrightarrow ~\exists \lambda _{i}>0,$ $u,v\in \mathbb{R}^{n}$%
\begin{eqnarray*}
u_{i}+v_{j} &\leq &\lambda _{i}R_{ij} \\
u_{i}+v_{i} &=&0
\end{eqnarray*}

$\Longleftrightarrow ~\exists \lambda _{i}>0,$ $v\in \mathbb{R}^{n}$%
\begin{equation*}
v_{j}-v_{i}\leq \lambda _{i}R_{ij},
\end{equation*}

which is (ii) in Theorem 0.
\end{proof}

\bigskip

The previous result leads to the following two remarks.

\bigskip

First, recall that matrix $\left( M_{ij}\right) $ is called \emph{cyclically
monotone} if for any cycle $\left( i_{1},...,i_{p}\right) $, one has%
\begin{equation*}
\sum_{k=1}^{p}M_{i_{k}i_{k+1}}-M_{i_{k}i_{k}}\geq 0.
\end{equation*}

It is well known (see e.g. Villani, 2003) that cyclical monotonicity is
equivalent to (\ref{optTransp}). Therefore, Theorem (\ref%
{ThmCharacterizationGARP}) states that $\left( R_{ij}\right) $ is cyclically
consistent if and only if there are weights $\lambda _{i}>0$ such that $%
\lambda _{i}R_{ij}$ is cyclically monotone.

\bigskip

As a second remark, introduce%
\begin{equation*}
\mathcal{F}=conv\left( \left( c_{i\sigma \left( i\right) }\right)
_{i=1,...,n}:\sigma \in S\right) \subset \mathbb{R}^{n}
\end{equation*}%
which is a convex polytope. Note that%
\begin{equation*}
\min_{\sigma \in S}\sum_{i=1}^{n}\lambda _{i}c_{i\sigma \left( i\right)
}=\min_{C\in \mathcal{F}}\sum_{i=1}^{n}\lambda _{i}C_{i}.
\end{equation*}

\bigskip

We have the following geometric characterization of the fact that the data
are rationalizable:

\bigskip

\begin{proposition}
The data are rationalizable if and only if $0$\ is an extreme point of $%
\mathcal{F}$\ with a componentwise positive vector in the normal cone.
\end{proposition}

\bigskip

\begin{proof}
Introduce $\mathcal{W}\left( \lambda \right) =\min_{C\in \mathcal{F}%
}\sum_{i=1}^{n}\lambda _{i}C_{i}$. This concave function is the support
function of $C$. $C_{i}^{0}=c_{i\sigma _{0}\left( i\right) }$ is an extreme point of
$\mathcal{F}$ with a componentwise positive vector in the normal cone if and
only $0$ is in the superdifferential of $\mathcal{W}$ at such a vector $%
\lambda $. This holds if and only if $\mathcal{W}\left( \lambda \right)
=\sum_{i=1}^{n}\lambda _{i}C_{i}^{0}$.
\end{proof}

\bigskip

\section{Strong and weak rationalizability}

\subsection{Indices of rationalizability}

Rationalizability of the data by a single consumer, as tested by Afriat's
inequalities, is an important empirical question. Hence it is of interest to
introduce measures of how close\ the data is from being rationalizable.
Since Afriat's original work on the topic, many authors have set out
proposals to achieve this. Afriat's original \textquotedblleft efficiency
index\textquotedblright\ is the largest $e\leq 1$ such that $%
R_{ij}^{e}=R_{ij}+\left( 1-e\right) b_{i}$ is cyclically consistent, where $%
b_{i}>0$ is a fixed vector with positive components. In Afriat's setting, $%
R_{ij}=x_{j}\cdot p_{i}-x_{i}\cdot p_{i}$, and $R_{ij}^{e}=x_{j}\cdot
p_{i}-ex_{i}\cdot p_{i}$, so $b_{i}=x_{i}\cdot p_{i}$. Many other tests and
empirical approaches have followed and are discussed in Varian's (2006)
review paper. A recent proposal is given in Echenique, Lee and Shum (2011),
who introduce the \textquotedblleft money pump index\textquotedblright\ as
the amount of money that could be extracted from a consumer with
non-rationalizable preferences.

In the same spirit, we shall introduce indices that will measure how far the
dataset is from being rationalizable. The indices we shall build are
connected to the dual interpretation of revealed preferences we have
outlined above. Our measures of departure from rationalizability will be
connected to measures of departure from Pareto efficiency in that problem.
One measure of departure from efficiency is the welfare gains that one would
gain from moving to the efficient frontier; this is interpretable as
Debreu's (1951) \emph{coefficient of resource utilization}\footnote{%
We thank Don Brown for suggesting this interpretation to us.}, in the case
of a convex economy. This is only an analogy, as we are here in an
indivisible setting, but this exactly the idea we shall base the
construction of our indices on.

\bigskip

It seems natural to introduce index $A$ as%
\begin{equation*}
A=\max_{\lambda \in \Delta }\min_{\sigma \in S}\sum_{i=1}^{n}\lambda
_{i}R_{i\sigma \left( i\right) }
\end{equation*}%
where $\Delta =\left\{ \lambda \geq 0,\sum_{i=1}^{n}\lambda _{i}=1\right\} $
is the simplex of $\mathbb{R}^{n}$. Indeed, we have%
\begin{equation*}
A\leq 0,
\end{equation*}%
and by compacity of $\Delta $, equality holds if and only if there exist $%
\lambda \in \Delta $ such that 
\begin{equation*}
\min_{\sigma \in S}\sum_{i=1}^{n}\lambda _{i}R_{i\sigma \left( i\right) }=0.
\end{equation*}

\bigskip

Of course, this differs from our characterization of rationalizability in
Theorem \ref{ThmCharacterizationGARP}, as there the weights $\lambda _{i}$'s
need to be all positive, not simply nonnegative. It is easy to construct
examples where (\ref{negishiOpt}) holds with some zero $\lambda _{i}$'s and $%
\sigma _{0}\left( i\right) =i$ is not efficient. For example, in the housing
problem, if individual $i=1$ has his most preferred option, then $\lambda
_{1}\neq 0$ and all the other $\lambda _{i}$'s are zero, and $A=0$, thus (%
\ref{negishiOpt}) holds. However, allocation may not be Pareto because there
may be inefficiencies among the rest of the individuals.

\bigskip

Hence imposing $\lambda >0$ is crucial. Fortunately, it turns out that one
can restrict the simplex $\Delta $ to a subset which is convex, compact and
away from zero, as shown in the next lemma.

\bigskip

\begin{lemma}
\label{compactification}There is $\varepsilon >0$ (dependent only on the
entries of matrix $c$, but independent on the matching considered) such that
the following disjuction holds

\begin{itemize}
\item either there exist no scalars $\lambda _{i}>0$ satisfying the
condtions in Theorem \ref{ThmCharacterizationGARP}

\item or there exist scalars $\lambda _{i}>0$ satisfying the condtions in
Theorem \ref{ThmCharacterizationGARP} and such that%
\begin{equation*}
\left\{ 
\begin{array}{c}
\lambda _{i}\geq \varepsilon \text{ for all }i, \\ 
\sum_{i=1}^{n}\lambda _{i}=1\text{.}%
\end{array}%
\right.
\end{equation*}
\end{itemize}
\end{lemma}

\begin{proof}
Standard construction (see \cite{fostel}) of the $\lambda _{i}$'s and the $%
v_{i}$'s provides a deterministic procedure that returns strictly positive $%
\lambda _{i}\geq 1$ within a finite and bounded number of steps, with only
the entries of $R_{ij}$ as input; hence $\lambda $, if it exists, is
bounded, so there exists $M$ depending only on $R$ such that $%
\sum_{i=1}^{n}\lambda _{i}\leq M$. \ Normalizing $\lambda $ so that $%
\sum_{i=1}^{n}\lambda _{i}=1$, one sees that one can choose $\varepsilon
=1/M $.
\end{proof}

\bigskip

We denote $\Delta _{\varepsilon }$ the set of such vectors $\lambda $, and
we recall that $\Delta $ is the set of $\lambda $ such that $\lambda
_{i}\geq 0$ for all $i$ and $\sum_{i=1}^{n}\lambda _{i}=1$. Recall $%
R_{ij}=c_{ij}-c_{ii}$, and introduce%
\begin{equation*}
A^{\ast }=\max_{\lambda \in \Delta _{\varepsilon }}\min_{\sigma \in
S}\sum_{i=1}^{n}\lambda _{i}R_{i\sigma \left( i\right) },
\end{equation*}%
so that we have%
\begin{equation*}
A^{\ast }\leq 0\text{ }
\end{equation*}%
and equality if and only if the data are rationalizable (as in this case
there exist $\lambda _{i}>0$ such that the characterization of
rationalizability in \ref{ThmCharacterizationGARP} is met). Further, as $%
\Delta _{\varepsilon }\subset \Delta $, one gets%
\begin{equation*}
\underset{A^{\ast }}{\underbrace{\max_{\lambda \in \Delta _{\varepsilon
}}\min_{\sigma \in S}\sum_{i=1}^{n}\lambda _{i}R_{i\sigma \left( i\right) }}~%
}\leq ~\underset{A}{\underbrace{\max_{\lambda \in \Delta }\min_{\sigma \in
S}\sum_{i=1}^{n}\lambda _{i}R_{i\sigma \left( i\right) }}}~\leq ~0.
\end{equation*}

\bigskip

The max-min formulation for index $A$ leads naturally to introduce a new
index which comes from the dual min-max program%
\begin{eqnarray*}
B &=&\min_{\sigma \in S}\max_{\lambda \in \Delta }\sum_{i=1}^{n}\lambda
_{i}R_{i\sigma \left( i\right) } \\
&=&\min_{\sigma \in S}\max_{i\in \left\{ 1,...,n\right\} }R_{i\sigma \left(
i\right) }
\end{eqnarray*}%
Note that the inequality $\max \min \leq \min \max $ always hold, so $A\leq
B $, and further, we have%
\begin{equation*}
B\leq \max_{i\in \left\{ 1,...,n\right\} }R_{ii}=0.
\end{equation*}%
Therefore we have%
\begin{equation*}
\underset{A^{\ast }}{\underbrace{\max_{\lambda \in \Delta _{\varepsilon
}}\min_{\sigma \in S}\sum_{i=1}^{n}\lambda _{i}R_{i\sigma \left( i\right) }}~%
}\leq ~\underset{A}{\underbrace{\max_{\lambda \in \Delta }\min_{\sigma \in
S}\sum_{i=1}^{n}\lambda _{i}R_{i\sigma \left( i\right) }}}~\leq ~\underset{B}%
{\underbrace{\min_{\sigma \in S}\max_{\lambda \in \Delta
}\sum_{i=1}^{n}\lambda _{i}R_{i\sigma \left( i\right) }}}~\leq ~0.
\end{equation*}

\bigskip

\bigskip

We shall come back to the interpretation of $A^{\ast }=0$, $A=0$ and $B=0$
as stronger or weaker forms of rationalizability of the data. Before we do
that, we summarize the above results.

\bigskip

\begin{proposition}
\label{PropEquivIndices}We have:

(i) $A^{\ast }=0$ if and only if there exist scalars $v_{i}$ and weights $%
\lambda _{i}>0$ such that%
\begin{equation*}
v_{j}-v_{i}\leq \lambda _{i}R_{ij}.
\end{equation*}

(ii) $A=0$ if and only if there exist scalars $v_{i}$ and weights $\lambda
_{i}\geq 0$, not all zero, such that%
\begin{equation*}
v_{j}-v_{i}\leq \lambda _{i}R_{ij}.
\end{equation*}

(iii) $B=0$ if and only if 
\begin{equation*}
\forall \sigma \in S,~\exists i\in \left\{ 1,...,n\right\} :R_{i\sigma
\left( i\right) }\geq 0.
\end{equation*}

(iv) One has%
\begin{equation}
A^{\ast }\leq A\leq B\leq 0.  \label{ineq}
\end{equation}
\end{proposition}

\bigskip

Note that it is easy to find examples where these inequalities hold strict
for $n\geq 3$.

\begin{proof}
(i) follows from Lemma \ref{compactification}. To see (ii), note that $A=0$
is equivalent to the existence of $\lambda \in \Delta $ such that $%
\min_{\sigma \in S}\sum_{i=1}^{n}\lambda _{i}R_{i\sigma \left( i\right) }=0$%
, and the rest follows from Theorem \ref{ThmCharacterizationGARP}. The condition in (iii) is
equivalent to the fact that for all $\sigma \in S$, $\max_{i\in \left\{
1,...,n\right\} }R_{i\sigma \left( i\right) }\geq 0$, that is for all $%
\sigma \in S$, there exists $i\in \left\{ 1,...,n\right\} $ such that $%
R_{i\sigma \left( i\right) }\geq 0$.\ The inequalities in (iv) were
explained above.
\end{proof}

\bigskip

\bigskip

In an unpublished manuscript (\cite{geanakoplos}) that he kindly
communicated to us on our request, John Geanakoplos introduces the following
minmax problem, which in our notations can be defined as%
\begin{equation*}
G=\max_{\lambda \in \Delta }\min_{\sigma \in C}\sum_{i=1}^{n}\lambda
_{i}R_{i\sigma \left( i\right) }
\end{equation*}%
where $C\subset S$ is the set of permutations that have only one cycle, ie.
such that there exist a cycle $i_{1},...,i_{p+1}=i_{1}$ such that $\sigma
\left( i_{k}\right) =i_{k+1}$ and $\sigma \left( i\right) =i$ if $i\notin
\left\{ i_{1},...,i_{p}\right\} $. As $C\subset S$ one has%
\begin{equation*}
A\leq G\leq 0.
\end{equation*}

Geanakoplos uses index $G$ and von Neuman's minmax theorem in order to
provide an interesting alternative proof of Afriat's theorem. However, it
seems that index $G$ is not directly connected with the assignment problem.

\bigskip

\subsection{Interpretation of indexes $A^{\ast }$, $A$, $B$}

As seen above, the index $A^{\ast }$ was constructed so that $A^{\ast }=0$
if and only if the data are rationalizable. The indexes $A$ and $B$ will
both be equal to $0$ if the data are rationalizable; hence $A<0$ or $B<0$
imply that the data is not rationalizable; however the converse is not true,
so these indexes can be interpreted as measures of weaker form of
rationalizability. Hence we would like to give a meaningful interpretation
of the situations where $A=0$ and $B=0$. It turns out that $A=0$ is
equivalent to the fact that a subset of the observations can be
rationalized, the subset having to have a property of \emph{coherence} that
we now define. $B=0$ is equivalent to the fact that one cannot partition the
set of observations into \emph{increasing cycles}, a notion we will now
define. In other words, the condition $B<0$ means that any observation is
part of a preference cycles--indeed a very strong departure from
rationalizability. These indexes may be used to compute identified regions
in models with partial identification; see Ekeland, Galichon and Henry
(2010) and references therein.

\bigskip

Throughout this subsection it will be assumed that no individual is
indifferent between two distinct consumptions for the direct revealed
preference relation, that is:

\bigskip

\textbf{Assumption A.} \emph{In this subsection, }$R$\emph{\ is assumed to
verify }$R_{ij}\neq 0$\emph{\ for }$i\neq j$\emph{.}

\bigskip

We first define the notion of coherent subset of observations.

\bigskip

\begin{definition}[Coherent subset]
\label{DefcoherentSubcoalitions}In the revealed preference problem, a subset
of observations included in $\left\{ 1,...,n\right\} $\ is said to be
coherent when $i\in I$\ and $i$\ directly revealed preferred to $j$\ implies 
$j\in I$. Namely, $I$ is coherent when%
\begin{equation*}
i\in I\text{ and }R_{ij}<0\text{ implies }j\in I.
\end{equation*}
\end{definition}

\bigskip

In particular, $\left\{ 1,...,n\right\} $ is coherent; any subset of
observations where each observation is directly revealed preferred to no
other one is also coherent. Next, we define the notion of increasing cycles.

\bigskip

\begin{definition}[Increasing cycles]
\label{DefIncreasingCycles}A cycle $i_{1},...,i_{p+1}=i_{1}$ is called \emph{%
increasing} when each observation is strictly directly revealed preferred to
its predecessor. Namely, cycle $i_{1},...,i_{p+1}=i_{1}$ is increasing when%
\begin{equation*}
R_{i_{k}i_{k+1}}<0\text{ for all }k\in \left\{ 1,...,p\right\} .
\end{equation*}
\end{definition}

\bigskip

Of course, the existence of an increasing cycle implies that the matrix $R$
is not cyclically consistent, hence it implies in the revealed preference
problem that the data are not rationalizable. It results from the definition
that an increasing cycle has length greater than one.

\bigskip

We now state the main result of this section, which provides an economic
interpretation for the indexes $A^{\ast }$, $A$ and $B$.

\bigskip

\begin{theorem}
\label{ThmcoherentSubcoalitions}We have:

(i) $A^{\ast }=0$\ iff the data are rationalizable,

(ii) $A=0$\ iff a coherent subset of the data is rationalizable,

(iii) $B=0$ iff there is no partition of $\left\{ 1,...,n\right\} $ in
increasing cycles.

and (i) implies (ii), which implies (iii).
\end{theorem}

\bigskip

\begin{proof}
$(i)$ was proved in Theorem \ref{ThmCharacterizationGARP} above.

\bigskip

Let us show the equivalence in $(ii)$. From Proposition \ref%
{PropEquivIndices} $A=0$ is equivalent to the existence of $\exists \lambda
_{i}\geq 0$, $\sum_{i=1}^{n}\lambda _{i}=1$ and $v\in \mathbb{R}^{n}$ such
that
\begin{equation*}
v_{j}-v_{i}\leq \lambda _{i}R_{ij},
\end{equation*}%
so defining $I$ as the set of $i\in \left\{ 1,...,n\right\} $ such that $%
\lambda _{i}>0$, this implies a subset $I$ of the data is rationalizable. We
now show that $I$ is coherent. Indeed, for any two $k$ and $l$ not in $I$
and $i$ in $I$, one has $v_{k}=v_{l^{\prime }}\geq v_{i}$; thus if $i\in I$
and $R_{ij}<0$, then $v_{j}<v_{i}$, hence $j\in I$, which show that $I$ is
coherent.

Conversely, assume a coherent subset of the data $I$ is rationalizable. Then
there exist $\left( u_{i}\right) _{i\in I}$ and $\left( \mu _{i}\right)
_{i\in I}$ such that $\mu _{i}>0$ and%
\begin{equation*}
u_{j}-u_{i}\leq \mu _{i}R_{ij}
\end{equation*}%
for $i,j\in I$. Complete by $u_{i}=\max_{k\in I}u_{k}$ for $i\notin I$, and
introduce $\tilde{R}_{ij}=1_{\left\{ i\in I\right\} }R_{ij}$. One has $%
\tilde{R}_{ij}<0$ implies $i\in I$ and $R_{ij}<0$ hence $j\in I$ by the
coherence property of $I$, thus $u_{j}-u_{i}<0$. Now $\tilde{R}_{ij}=0$
implies either $i\notin I$ or $R_{ij}=0$ thus $i=j$; in both cases, $%
u_{j}\leq u_{i}$. Therefore, one has%
\begin{eqnarray*}
\tilde{R}_{ij} &<&0\text{ implies }u_{j}-u_{i}<0\text{, and} \\
\tilde{R}_{ij} &\leq &0\text{ implies }u_{j}-u_{i}\leq 0\text{.}
\end{eqnarray*}

Hence by Theorem 0, there exist $v_{i}$ and $\bar{\lambda}_{i}>0$ such that%
\begin{equation*}
v_{j}-v_{i}\leq \bar{\lambda}_{i}\tilde{R}_{ij}
\end{equation*}%
and defining $\lambda _{i}=\bar{\lambda}_{i}1_{\left\{ i\in I\right\} }$,
one has
\begin{equation*}
v_{j}-v_{i}\leq \lambda _{i}R_{ij}
\end{equation*}

which is equivalent to $A=0$.

\bigskip

$(iii)$ Now Proposition \ref{PropEquivIndices} implies that $B<0$ implies
that there is $\sigma \in S$ such that $\forall i\in \left\{ 1,...,n\right\}
,~R_{i\sigma \left( i\right) }<0$. The decomposition of $\sigma $ in cycles
gives a partition of $\left\{ 1,...,n\right\} $ in increasing cycles.

\bigskip

$(iv)$ The implication $(i)\Rightarrow (ii)\Rightarrow (iii)$ results from
inequality $A^{\ast }\leq A\leq B\leq 0$.\bigskip
\end{proof}

\section{Concluding remarks}

To conclude, we make a series or remarks.

\bigskip

First, we have shown how our dual interpretation of Afriat's theorem in its
original Revealed Preference context as well as in a less traditional
interpretation of efficiency in the housing problem could shed new light and
give new tools for both problems. It would be interesting to undestand if
there is a similar duality for problems of revealed preferences in matching
markets, recently investigated by Galichon and Salani\'{e} (2010) and
Echenique et al. (2011).

\bigskip

Also, we argue that it makes sense to investigate \textquotedblleft weak
rationalizability\textquotedblright\ (ie. $A=0$ or $B=0$) instead of
\textquotedblleft strong rationalizability\textquotedblright\ (ie. $A^{\ast
}=0$), or equivalently, it may make sense to allow some $\lambda _{i}$'s to
be zero. In the case of $A=0$, recall the $\lambda _{i}$'s are interpreted
in Afriat's theory as the Lagrange multiplier of the budget constraints.
Allowing for $\lambda =0$ corresponds to excluding wealthiest individuals as
outliers. It is well-known that when taken to the data, strong
rationalizability is most often rejected. It would be interesting to test
econometrically for weak rationalizability, namely whether $A=0$.

$B<0$ is a very strong measure of nonrationalizability, as $B<0$ means that
one can find a partition of the observation set into increasing cycles,
which seems a very strong violations of the Generalized Axiom of Revealed
Preference. But this may arise in some cases, especially with a limited
number of observations.

\bigskip

We should emphasize on the fact that indexes $A^{\ast }$, $A$ and $B$
provide a measure of how close the data is from being rationalizable, in the
spirit of Afriat's efficiency index. It is clear for every empirical
researcher that the relevant question about revealed preference in consumer
demand is not whether the data satisfy GARP, it is how much they violate it.
These indexes are an answer to that question. Also the geometric
interpretation of rationalizable is likely to provide useful insights for
handling unobserved heterogeneity. We plan to investigate this question in
further research.

\bigskip 

Last, it seems that an interesting research avenue (not pursued in the
present paper) deals with exploring similar connections as in this paper,
and understand what would be the analogue of variants of the theories
discussed here. For instance, in the light of the duality exposed in the
present paper, it would be interesting to characterize the dual theory to
the Revealed Preference theory for collective models of household
consumption as in Cherchye, De Rock and Vermeulen (2007, 2010). Similarly,
it would also be of interest to understand the Revealed Preference dual of
the \emph{two-sided} version of the housing problem, namely the marriage
problem, where both sides of the market have preferences over the other
side. In these two cases as in others, what becomes of the connection
stressed in the current paper? this question, too, is left for future
research.

\end{document}